\documentclass[12pt]{article}
\usepackage[textures]{graphics}

\parskip=\bigskipamount
\newcommand{\beq}{\begin{equation}}
\newcommand{\eeq}{\end{equation}}
\newcommand{\beqa}{\begin{eqnarray}}
\newcommand{\eeqa}{\end{eqnarray}}
\newcommand{\ket} [1] {\vert #1 \rangle}
\newcommand{\bra} [1] {\langle #1 \vert}
\newcommand{\braket}[2]{\langle #1 | #2 \rangle}

\newcommand{\mod}{~{\rm mod}~}
\newcommand{\red}{\mbox{\rm \footnotesize red}}

\begin{document}
\title{{ Security of Quantum Key Distribution with entangled qu$N$its.}}
\author{Thomas Durt$^1$, Dagomir Kaszlikowski$^2$, Jing-Ling Chen$^2$, L.C. Kwek$^2$$^-$$^3$,\\
 $^1$ Toegepaste Natuurkunde en Fotonica, Theoretische Natuurkunde,\\ 
Vrije Universiteit Brussel, Pleinlaan 2, 1050 Brussels, Belgium\\
$^2$Department of Physics, Faculty of Science,\\
 National University of Singapore, Lower Kent Ridge, Singapore 119260, \\
$^3$National Institute of Education, Nanyang Technological University,\\
 1 Nanyang Walk, Singapore 639798}
 \date{}

\maketitle

\begin{abstract}
We consider a generalisation of Ekert's entanglement-based quantum
cryptographic protocol where qubits are replaced by qu$N$its
(i.e., $N$-dimensional systems). In order to study its robustness
against optimal incoherent attacks, we derive the information
gained by a potential eavesdropper during a cloning-based
individual attack. In doing so, we generalize Cerf's formalism
for cloning machines and establish the form of the most general
cloning machine that respects all the symmetries of the problem.
We obtain an upper bound on the error rate that guarantees the
confidentiality of qu$N$it generalisations of the Ekert's protocol
for qubits.
\end{abstract}
\maketitle
 
\section{Introduction}
 
In quantum cryptographic protocols, the presence of an
eavesdropper in the communication channel can be revealed through
disturbances in the transmission of the message. To realize such
protocols, it is necessary to encode the signal into quantum
states that belong to non-compatible bases, as in the original
protocol of Bennett and Brassard\cite{BB84}. In 1991, Ekert
suggested\cite{Ekert} a scheme in which the security of quantum
cryptography is based on entanglement. In this scheme,  one
encrypts the key into the non-compatible qubit bases that
maximize the violation of local realism.
 
Recently it was shown that this violation is more pronounced for
the case of entangled qu$N$its\cite{zukozeil,Durt,collins} for $N
> 2$. Moreover, the qutrit generalisation of Ekert's protocol is
more robust and safer than its qubit
counterpart\cite{dagomir,Durttrits}. This naturally serves as a
significant motivation for studying the generalisation of Ekert's
protocol where qubits are replaced by qu$N$its.
 
There are several ways of realizing qu$N$its experimentally. One
realization of qu$N$its, possibly the most straightforward one,
exploits time-bins\cite{Tittel}. This approach has already been
demonstrated for entangled photons up to eleven
dimensions\cite{DeRiedmatten}. Another possibility is to utilize
multiport-beamsplitters, and more specifically those that split
the incoming single light beam into $N$\cite{ZZH} outputs. Thus,
an entanglement-based quantum cryptographic protocol based on
qu$N$its instead of qubits can be realized with the current
state-of-the-art quantum optical techniques.
 
In this paper, we establish the generality of a class of
eavesdropping attacks that are based on (state-dependent) quantum
cloning machines\cite{CERFPRL,CERFA,CERF,buzek,werner,DBruss}. We
shall show that such attacks cannot be thwarted by Alice and Bob,
the authorized users of the quantum cryptographic channel,
because the disturbance due to the presence of the eavesdropper
(Eve) perfectly mimics the correlations of an unbiased noise, at
least for what concerns correlations between the encryption and
decryption bases that characterize the $N$ dimensional Ekert
protocol.  The security is shown to be higher for higher
dimensional systems, a property that was already noted for
several qutrit-based protocols in comparison to their qubit-based
counterparts\cite{dagomir,
Durttrits,boure1,HBech-APeres,bourennane,bruss3D}.

\section{The four qu$N$it bases that maximize the violation of local realism}
 
In the Ekert91 protocol\cite{Ekert}, the four qubit bases chosen
by Alice and Bob are the bases that maximize the violation of
CHSH inequalities\cite{chsh}. There exists a natural
generalisation of this set of bases in the case of
qu$N$its\cite{Durt}. In analogy with the Ekert91 bases that
belong to a great circle on the Bloch sphere, these bases belong
to a set of bases parametrized by a phase $\phi$, that we shall from now on call the $\phi$ bases. These bases are
related to the computational basis $\{\ket{0},\ket{1},...,\ket{N-1}\}$:
 
\beq \label{optimal}\ket{l_{\phi}}={1\over \sqrt
N}\Sigma_{k=0}^{N-1} e^{ik({2\pi\over N}l+\phi)}\ket{k},
l=0,...,N-1\eeq
 
It has been shown that when local observers measure the
correlations exhibited by the maximally entangled state
$\ket{\phi_N^+}={1\over\sqrt{N}}\sum_{i=1}^{N-1}\ket{i}\otimes
\ket{i}$ in the four $\phi$ bases that we obtain when
$\phi_i\,=\,{2\pi\over 4N}\cdot i (i=0,1,2,3)$, the degree of
nonclassicality or violation of local realism that characterizes
the correlations increases with the dimension
$N$\cite{zukozeil,Durt,collins}. It is also higher than the
degree of nonclassicality allowed by Cirelson's
theorem\cite{cirelson} for qubits, and higher than for a large
class of other qu$N$it bases. Indeed, this can be shown by
estimating the resistance of non-locality against
noise\cite{zukozeil,Durt}, or by considering generalisations of
the CHSH inequality to bipartite entangled qu$N$it
states\cite{collins}. From now on, we call the four qu$N$it bases
that maximize the violation of local realism the optimal bases.

\section{The $N$-dimensional Entanglement based ($N$-DEB) protocol}

Let us now assume that the source emits the maximally entangled
qu$N$it state $\ket{\phi_N^+}$ and that Alice and Bob share this
entangled pair and perform measurements along one of the optimal
bases described in the previous section.
 
In order to show that these bases are pairwise perfectly
correlated, it is useful to present an interesting property of
the state $\ket{\phi_N^+}$. Let us consider two bases: the $\psi$
basis chosen arbitrarily (with $\braket{i}{\psi_j}$ = $ U_{ij}$),
and its conjugate basis, the $\psi^*$ basis (with
$\braket{i}{\psi^*_j}$ = $ U^*_{ ij}$). When Alice and Bob share
the maximally entangled state $\ket{\phi_N^+}$ and that Alice
measures it in the $\psi^*$ basis and Bob in the $\psi$ basis,
their results are  perfectly correlated. To see this, we note
that by virtue of the unitarity of the matrix $U_{ij}$,
$\ket{\phi_N^+}={1\over\sqrt{N}} \sum_{k,l,m=0}^{N-1}$
$\ket{\psi^*_l}\braket{\psi^*_l}{k}\otimes$
 $\ket{\psi_m}\braket{\psi_m}{k}$ =$ {1\over\sqrt{N}} \sum_{k,l,m=0}^{N-1}
\ket{\psi^*_l} U_{ kl}\ket{\psi_m} U^*_{ km}$ = ${1\over\sqrt{N}}
\sum_{l,m=0}^{N-1} \ket{\psi^*_l} \ket{\psi_m}\delta_{ ml}$ =
${1\over\sqrt{N}} \sum_{k=0}^{N-1} \ket{\psi^*_k}\ket{\psi_k}$,
whenever Alice projects the state $\ket{\phi_N^+}$ into the
conjugate basis $\psi^*$, she projects Bob's component into the
$\psi$ basis and reciprocally.

Moreover, the state $\ket{\phi_N^+}$ can be rewritten as
$\ket{\phi_N^+}={1\over\sqrt{N}}(\sum_{i=1}^{N-1}
(\ket{i_{\phi}^*}\otimes \ket{i_{\phi}})$ where \beq
\label{optimalconj}\ket{l_{\phi}^*}={1\over \sqrt
N}\Sigma_{k=0}^{N-1} e^{-ik({2\pi\over N}l+\phi)}\ket{k}
(l:0,...,N-1)\eeq
Therefore, when Bob performs a measurement in the $\phi$ basis
($\ket{k_{\phi}}$) and Alice in its conjugate basis
($\ket{k_{\phi}^*}$), their results are perfectly correlated. Now,
the four optimal bases are pairwise conjugate and hence perfectly
correlated as well. To show the perfect correlation, we note that
for the phase that appears in Eq.~(\ref{optimalconj}),
$-k({2\pi\over N}l+\phi)= k({2\pi\over N}(N-l-j)-\phi +
j{2\pi\over N})$ $ \mod 2\pi$ where $j$ is an arbitrary integer
number. Now, $N-l-j$ varies from 0 to $N-1$ ($\mod N$) when $l$
varies from 0 to $N-1$ which shows that the $\phi^*$ basis is the
same as the $\phi'$ basis (with $\phi'$=$-\phi + j{2\pi\over
N}$). It is easy to check that, thanks to an appropriate choice
of $j$, the bases associated to even values of $i$
$(i=0,2,\phi_i\,=\,{2\pi\over 4N}\cdot i)$ are preserved under
phase conjugation, while the bases associated to odd values of
$i$ $(i=1,3)$ are interchanged. The four optimal bases are thus
pairwise maximally correlated.

In a natural generalisation of the Ekert91 protocol for qu$N$its,
denoted as the $N$-DEB protocol (i.e. the $N$-dimensional
entanglement based protocol) in analogy with the notation adopted
in Ref.\cite{Durttrits} for the case $N = 3$, Alice and Bob share
the entangled state $\ket{\phi_N^+}$ and choose their measurement
basis at random among one of the four optimal qu$N$it bases
(according to the statistical distribution that they consider to
be optimal). Because of the existence of perfect correlations
between the conjugate bases, a fraction of the measurement
results can be used in order to establish a deterministic
cryptographic key. The rest of the data, for which Alice's and
Bob's bases are not perfectly correlated, can be used in
principle in order to detect the presence of an eavesdropper for
example with the help of generalized CHSH
inequalities\cite{collins}. Let us now study the safety of this
protocol against optimal incoherent attacks.

\section{Individual attacks and optimal cloning
machines}
 
We assume from now on that the noise that characterizes the
transmission (this includes dark counts in the detectors,
misalignments of Alice and Bob's bases, transmission losses and
so on) is unbiased (this is a very general assumption). Under
such conditions, the state shared by Alice and Bob must mimic the
correlations that is observed when they share a state that has
the following form:
\begin{eqnarray} &\rho_{R,A}(F_{N})=
(1-F_{N})\ket{\phi_N^+}_{R,A}\bra{\phi_N^+}_{R,A}
+F_{N}\rho^{noise}_{R,A},& \label{werner}
\end{eqnarray}
where $\rho^{noise}_{R,A}=\frac{1}{N^2}\hat{I}_{R,A}$, and the
positive parameter $F_{N}\leq1$ determines the ``noise fraction"
within  the full state. In the following, we make the
conservative hypothesis that all the errors of transmission could
be due to the presence of an eavesdropper who possesses a perfect
(noiseless) technology, controls the line of transmission and
lets her state(s) $\ket{\phi_N^+}_{R,A}$ interact with the one
originally shared by Alice and Bob and an auxiliary system (or
probe). In principle, Eve is free to use any interaction between
one or several qu$N$it pairs (originally prepared in the state
$\ket{\phi_N^+}_{R,A}$) and an auxiliary system of her choice.
Eve could keep this auxiliary system isolated and unperturbed
during an arbitrary long period of time. On hearing the public
discussion between Alice and Bob, she performs the measurement of
her choice on her system. In principle she could let her
auxiliary system interact with a series of successive signals
and/or carry out her measurement on a series of auxiliary
systems. We limit ourselves to the situation where she couples
each signal individually to an auxiliary system (individual or
incoherent attacks). It could happen that coherent attacks
(so-called joint or collective attacks, for a review, see e.g.
\cite{RMP}) are more dangerous but presently nobody knows whether
this is the case.
 
Let us assume that initially Alice and Bob share the state
$\ket{\phi_N^+}$ and that during an individual attack, Eve lets
this state interact with her probe (which also belongs to a
$N^2$-dimensional Hilbert space\cite{RMP} which can be seen as
the ``mirror space'' of Alice and Bob's $N^2$-dimensional Hilbert
space). Then, the most general cloning state
$\ket{\Psi}_{R,A,B,C}$ where the labels $R, A,B,C$ are
respectively associated to the reference $R$ (Alice), the two
output clones ($A$ for Bob and $B$ for Eve), and to the
($N$-dimensional) cloning machine ($C$) is the element of a $N^4$
dimensional Hilbert space. Our task is to optimize this state in
such a way that Eve maximizes her information and minimizes the
disturbance on the information exchanged between Alice and Bob.
Moreover, as the real disturbance along a transmission line is
assumed to be unbiased, the disturbance induced by the presence
of Eve must mimic an isotropic disturbance so to say, it may not
depend (1) on the state that Alice and Bob measure when they
measure in perfectly correlated (conjugate) bases, and also (2) on
which pair of such bases is selected; finally (3) it must also
mimic the correlations between the non-conjugate bases that would
be observed in the presence of real (unbiased) noise.

\subsection{Invariance under relabelings of the detectors and Cerf states}
 
Clearly the complexity of the problem to solve increases as the
fourth power of the dimension $N$. It is thus necessary to
simplify the treatment as much as possible  by taking account of
the intrinsic symmetries of the problem. According to the
notation introduced in the previous section, let us consider an
arbitrary basis, the $\phi$ basis associated to Bob, and its
conjugate basis, the $\phi^*$ basis, associated to Alice. A
fundamental symmetry characterizes such bases: the states of the
$\phi$ ($\phi^*$) basis are permuted if we vary the phase $\phi$
by any multiple of ${2\pi\over 3}$. It is easy to show that any
such permutation is generated by $C$, the generator of the cyclic
permutations that shifts each label of the states of the $\phi$
($\phi^*$) basis by unity ($l \to l+1$ ($\mod N$)). When the
condition (1) is fulfilled, so to say when the error rate does
not depend on the label $l$, it is natural to impose this
invariance at the level of the cloning state. In order to do so,
it is useful to introduce the so-called Bell states.  The $N^2$
generalized Bell states are defined as follows\cite{Durttrits}:
\beq \label{def1} \ket{B^{\psi}_{m^*,n}}_{R,A}=N^{-1/2}
\sum_{k=0}^{N-1} e^{2\pi i (kn/N)}
\ket{\psi_k^{*}}_{R}\ket{\psi_{k+m}}_{A} \eeq with $m$ and $n$
($0\le m,n \le N-1$) labeling these Bell states. Obviously they
are eigenstates under the generator of cyclic permutations $l \to
l+1$ for the eigenvalue $e^{-2\pi i (n/N)}$. Moreover they form
an orthonormal basis of the $N^2$ dimensional space spanned by
the product states $\ket{\psi_k^{*}}_{R}\ket{\psi_{l}}_{A}$
$(l,j=0,...,N-1)$.
 
Note that all the Bell states are maximally entangled and that
$\ket{\phi_N^+}=\ket{B^{\psi}_{0,0}}$. Let us consider the
fraction of the signal that is measured by Alice in the $\psi$
basis and by Bob in its conjugate basis (the $\psi^*$ basis). As
these are perfectly correlated bases, this signal is not discarded
and it will be used afterwards, during the reconciliation
protocol, to establish a fresh cryptographic key. At this level,
when Alice and Bob reveal publicly their choices of bases, Eve
will measure her ancilla in the basis that she deems to be
optimal\cite{RMP}. Eve is free to redefine this basis thanks to
an arbitrary unitary transformation according to her convenience.
Taking account of the dual nature of the ``mirror'' space
assigned to Eve, it is natural that Eve fixes this unitary
transformation in such a way that she measures her copy in the
same basis as Bob (the $\psi$ basis) and the cloning state in the
conjugate basis (the $\psi^*$ basis). Expressed in these bases,
the most general state $\ket{\Psi}_{R,A,B,C}$ that is invariant
under cyclic permutations of the labels assigned to the detectors
has then (up to an arbitrary redefinition of Eve's basis) the
following form: \beq \label{rho} \ket{\Psi}_{R,A,B,C}=
\sum_{m,m',n=1}^{N-1}a_{m,m',n}\ket{B^{\psi}_{m^*,n}}_{R,A}\ket{B^{\psi}_{m',-n^*}}_{B,C}\eeq
where by definition \beq \label{def2}
\ket{B^{\psi}_{m',-n^*}}_{B,C}=N^{-1/2} \sum_{k=0}^{N-1} e^{-2\pi
i (kn/N)}\ket{\psi_k}_{B}\ket{\psi^{*}_{k+m}}_{C} \eeq Note that
we introduced at this level two apparently different definitions
of the Bell states. Both can be re-expressed according to the
synthetic expression: \beq \label{def2} \ket{
B^{\psi}_{m^{(*)},n^{(*)}}}=N^{-1/2} \sum_{k=0}^{N-1} e^{2\pi i
(kn/N)} \ket{\psi_k^{(*)}}\ket{\psi_{k+m}^{(*)}} \eeq In the
computational basis (where $\ket{k}=\ket{k^*}$), these
definitions coincide with the usual definition\cite{CERF}.
Indeed, our approach constitutes a covariant generalisation of
Cerf's formalism for cloning machines\cite{CERF}. At this level,
the complexity of the problem to solve is only in $N^3$, because
we projected the state $\ket{\Psi}_{R,A,B,C}$ onto the set of
states that remains invariant under a shift unity of the labels
assigned to the basis states. We shall now show that it is still
possible to reduce the complexity of the problem if we consider
the question of optimality. Prior to this, let us introduce the
following definitions:

\noindent{\it Definition 1:}
The pure state $\ket{\Psi}_{R,A,B,C}$ is a Cerf state iff, for a
given basis (say the $\psi$ basis), \beq
\label{CERF}\ket{\Psi}_{R,A,B,C}=
\sum_{m,n=1}^{N-1}a_{m,n}\ket{B^{\psi}_{m^*,n}}_{R,A}\ket{B^{\psi}_{m,-n^*}}_{B,C}\eeq
 
Note that then $Tr_{B,C}\ket{\Psi}_{R,A,B,C}\bra{\Psi}_{R,A,B,C}$
is diagonal in the Bell basis $\ket{B^{\psi}_{m^*,n}}_{R,A}$ and
$Tr_{R,A}\ket{\Psi}_{R,A,B,C}\bra{\Psi}_{R,A,B,C}$ is diagonal in
the Bell basis $\ket{B^{\psi}_{m,n^*}}_{B,C}$. Beside,
normalization imposes that $\sum_{m,n}^{N-1}|a_{m,n}|^2 = 1$.
 
\noindent{\it Definition 2:}
 
A (pure) Cerf state is optimal among the Cerf states (for a given
quantum cryptographic protocol) iff, for the same mutual
information between Alice and Bob, the mutual information between
Alice and Eve corresponding to this state is superior or equal to
the one corresponding to any other (pure) Cerf state or to any
mixture of them.
 
\noindent{ \bf Theorem}:
 
-Let us assume that an attack is characterized by a state
$\ket{\Psi}_{R,A,B,C}$ that is invariant under cyclic
permutations of the labels of Alice and Bob's basis states (in
the $\psi^*$ and $\psi$ bases).
 
-If the optimal Cerf state exists, then it is also optimal among
all possible states $\ket{\Psi}_{R,A,B,C}$.
 
Proof:

We have shown that the most general state $\ket{\Psi}_{R,A,B,C}$
that is invariant under cyclic permutations of the labels of the
optimal bases must necessarily fulfill Eq.~(\ref{rho}) which does
not necessarily imply that it is a Cerf state. Nevertheless, by
virtue of Eq.~(\ref{def2}), we have that

\beqa\label{density}& \ket{\Psi}_{R,A,B,C}\bra{\Psi}_{R,A,B,C}=
N^{-2}\sum_{m,m',n,\tilde{m},\tilde{m'},
\tilde{n},k,l,\tilde{k},\tilde{l}=0}^{N-1}a_{m,m',n}a^*_{\tilde{m},\tilde{m'},\tilde{n}}
e^{i(2\pi/N)
((k-l).n-(\tilde{k}-\tilde{l}).\tilde{n})}. &\\
& \ket{\psi^{*}_{k}}_{R}\ket{\psi_{k+m}}_{A}
\ket{\psi_{l}}_{B}\ket{\psi_{l+m'}^*}_{C}
\bra{\psi^{*}_{\tilde{k}}}_{R}\bra{\psi_{\tilde{k}+\tilde{m}}}_{A}
\bra{\psi_{\tilde{l}}}_{B} \bra{ \psi_{\tilde l+\tilde m '}^*
}_{C} &\nonumber \eeqa

The mutual informations must be estimated on the non-discarded
signal that is measured in the $\psi^*$ basis by Alice and in the
$\psi$ basis by Bob, while Eve measures product states of the
type $\ket{\psi_{l}}_{B}\ket{\psi_{l'}^*}_{C}$. Therefore only
the diagonal coefficients that appear in the expression of the
density matrix (Eq.~(\ref{density})) are relevant. It is easy to
check that for such coefficients $m-m'=\tilde{m}-\tilde{m'}$, so
that $\ket{\Psi}_{R,A,B,C}\bra{\Psi}_{R,A,B,C}$ is equivalent to a
reduced density matrix $\rho^{\red}_{R,A,B,C}$ defined as follows:
$\rho^{\red}_{R,A,B,C}$ =
$\sum_{m,i,n,\tilde{m},\tilde{n}=0}^{N-1}a_{m,m'=m+i,n}a^*_{\tilde{m},\tilde{m'}
=\tilde{m}+i,\tilde{n}}$
$\ket{B^{\psi}_{m^*,n}}_{R,A}\ket{B^{\psi}_{m+i,-n^*}}_{B,C}
\bra{B^{\psi}_{\tilde{m}^*,\tilde{n}}}_{R,A}\bra{B^{\psi}_{\tilde{m}+i,-\tilde{n}^*}}_{B,C}$
which in turn corresponds to the following mixture:

$\rho^{\red}_{R,A,B,C}$ = $\sum_{i=0}^{N-1} P_i \rho^{i ~
\red}_{R,A,B,C}$, where \beq P_i \rho^{i ~ \red}_{R,A,B,C} =
\sum_{m,n=0}^{N-1}a_{m,m'=m+i,n}\ket{B^{\psi}_{m^*,n}}_{R,A}\ket{B^{\psi}_{m+i,-n^*}}_{B,C}
.\sum_{\tilde{m},\tilde{n}=0}^{N-1}a^*_{\tilde{m},\tilde{m'}=\tilde{m}+i,\tilde{n}}\bra{B^{\psi}_{\tilde{m}^*,
\tilde{n}}}_{R,A}\bra{B^{\psi}_{\tilde{m}+i,-\tilde{n}^*}}_{B,C}\nonumber\eeq
and $P_i=\sum_{m,n=0}^{N-1}|a_{m,m'=m+i,n}|^2$. $\rho^{i ~
\red}_{R,A,B,C}$ is the projector onto the state
$\ket{\Psi}_{R,A,B,C}^i$, with \beq\ket{\Psi}_{R,A,B,C}^i={1\over
\sqrt{P_i}}\sum_{m,n=0}^{N-1}a_{m,m'=m+i,n}
\ket{B^{\psi}_{m^*,n}}_{R,A}\ket{B^{\psi}_{m+i,-n^*}}_{B,C}.\nonumber\eeq

Everything happens as if the above state was chosen with
probability $P_i$ without that Eve is able to control this choice
or even to get informed about it. Her information is thus
certainly less than the information that she would get if she was
informed about this choice. Beside, once the choice of a
particular $\ket{\Psi}_{R,A,B,C}^i$ is realized, the mutual
information between Eve and Alice is invariant when Eve chooses
to re-label her detectors, in particular if she re-labels them
according to the rule $\ket{\psi_l}_{B}\ket{\psi_{l+m+i}^*}_{C}\to
\ket{\psi_l}_{B}\ket{\psi_{l+m}^*}$ which sends
$\ket{B^{\psi}_{m+i,-n^*}}_{B,C}$ onto
$\ket{B^{\psi}_{m,-n^*}}_{B,C}$. Note that this re-labeling does
not influence at all the statistical distribution of Alice and
Bob's results and their mutual information. In conclusion, Eve's
information is in the best case equivalent to the information
that she would get by realizing the state
$\ket{\Psi}_{R,A,B,C}^i={1\over
\sqrt{P_i}}\sum_{m,n=0}^{N-1}a_{m,m'=m+i,n}\ket{B^{\psi}_{m^*,n}}_{R,A}\ket{B^{\psi}_{m,-n^*}}_{B,C}$
with probability $P_i$, and being informed about the nature of
this choice. This corresponds to a mixture of Cerf states, which
ends the proof: when optimal Cerf state exist(s), then, if a state
$\ket{\Psi}_{R,A,B,C}$ belongs to the class of states defined in
Eq.~(\ref{rho}) and is optimal it is necessarily equal to this
(one of these) Cerf state(s).
 
This theorem shows that it is sufficient to optimize the cloning
machines described by a Cerf state which again reduces the
complexity of the problem: such a state is now described,
according to Eq.~(\ref{CERF}) by $N^2$ parameters $a_{m,n}$
instead of the $N^3$ parameters $a_{m,m',n}$. Note that in Cerf's
approach, Eq.~(\ref{CERF}) was considered to be an
ansatz\cite{CERF}; the previous theorem shows that its generality
can be established on the basis of more general assumptions. The
property that was encapsulated in the previous theorem expresses
a deep property of the Bell states. Actually, from Eve's
perspective, everything happens as if different families of Bell states
were separated by a classical super-selection rule. It helps to
understand why, when the optimal state is pure it is sufficient
to limit oneself to the quest of the optimal Cerf state.
 
\subsection{Invariance under the choice of the optimal basis}

At this level we did not exploit all the symmetries of the
problem, we only made use of the fact that inside a given pair of
perfectly correlated (conjugate) bases, the labeling of the
detectors is defined up to a cyclic permutation. Another symmetry
of the problem that we did not exploit yet is the following:
according to the condition (2), all pairs of perfectly correlated
bases must also be treated on equal footing. Therefore it is
natural to impose that the optimal Cerf state associated to the
$N$-DEB protocol fulfills the following constraints

\beqa\ket{\Psi}_{R,A,B,C}=
\sum_{m,n=1}^{N-1}a_{m,n}\ket{B^{\phi=0}_{m^*,n}}_{R,A}\ket{B^{\phi=0}_{m,-n^*}}_{B,C}=
\sum_{m,n=1}^{N-1}a_{m,n}\ket{B^{\phi={2\pi\over 4N}}_{m^*,n}}_{R,A}\ket{B^{\phi={2\pi\over 4N}}_{m,-n^*}}_{B,C}=\\
\sum_{m,n=1}^{N-1}a_{m,n}\ket{B^{\phi={4\pi\over
4N}}_{m^*,n}}_{R,A}\ket{B^{\phi={4\pi\over 4N}}_{m,-n^*}}_{B,C}=
\sum_{m,n=1}^{N-1}a_{m,n}\ket{B^{\phi={6\pi\over
4N}}_{m^*,n}}_{R,A}\ket{B^{\phi={6\pi\over
4N}}_{m,-n^*}}_{B,C}\eeqa

The treatment of this type of constraint is developed in
appendix. The result is extremely simple: whenever
$\braket{B^{\phi={2\pi p\over 4N}}_{i^*,j}}{ B^{\phi={2\pi q\over
4N}}_{k^*,l}}$ $\not = 0$ (where $p,q,=0,1,2,3)$, then $a_{i,j}$ =
$a_{k,l}$. Actually these results were already used in
\cite{Durttrits} for the treatment of the qutrit case but have
not been published yet. These constraints express the necessary
and sufficient conditions for which the Cerf state
(Eq.~(\ref{CERF})) that characterizes the cloning machine
possesses biorthogonal Schmidt decompositions in the Bell bases
(Eq.~(\ref{def2})) associated to the four optimal bases
simultaneously. By a straightforward computation we get that
\begin{eqnarray}
&&\braket{ B^{\phi_1}_{i^*,j} } { B^{\phi_2}_{k^*,l} } =
\braket{ B^{0}_{i^*,j} }{ B^{\phi_2-\phi_1}_{k^*,l} }=\nonumber\\
&&{1\over N}
\delta_{j,l}\sum_{p,q=0}^{N-1} \delta_{p-q,j (mod N)}\nonumber\\
&&e^{i(-p(\phi_2-\phi_1) + q((\phi_2-\phi_1)+ {2\pi \over
N}(k-i)))}
\end{eqnarray}

In particular, we have that when $j=l=0$,
$\braket{B^{\phi_1}_{i^*,0}}{ B^{\phi_2}_{k^*,0}}=\delta_{i,k}$,
which means that $B^{\phi=0}_{i^*,0}=B^{\phi}_{i^*,0}\forall
\phi$. When $j=l\not= 0$, $|\braket{B^{\phi_1}_{i^*,j}}{
B^{\phi_2}_{k^*,j}}|$ reaches the extremal values 1 or 0 only
when $\phi_1 - \phi_2 $ is an integer multiple of ${2\pi \over
N}$. This is due to the fact that for such values the basis
states $\ket{l_{\phi_1}}$ are equivalent to the states
$\ket{l_{\phi_2}}$, up to a cyclic permutation of the labels of
the basis states and we showed in a previous section that the
Bell states are eigenstates under such permutations. Otherwise,
for intermediate values of $\phi$ these in-products are never
equal to zero. Therefore the Cerf state $\ket{\Psi}_{R,A,B,C}$
that is invariant for at least two distinct values of $\phi$
($\mod {2\pi \over N}$) is characterized by the NxN matrix
$a_{m,n}$ that obeys the following equations:
\begin{equation}\label{asym1}
(a_{m,n})= \left(\begin{array}{ccccc}
v & y_1 & y_2 &...&y_{N-1}\\
x_1 & y_1 & y_2 & ...&y_{N-1}\\
x_2 & y_1  & y_2&...&y_{N-1} \\
. & .  & .&...&. \\
. & .  & .&...&. \\
. & .  & .&...&. \\
x_{N-1} & y_1  & y_2&...&y_{N-1} \\
\end{array}\right)
\end{equation}
The (normalized) matrix $a_{m,n}$ still contains $2(N-1)$
independent parameters. This shows that it is not enough to
impose the invariance of the state $\ket{\Psi}_{R,A,B,C}$ under
cyclic permutations of the basis states or under changes of bases
in order to fix all the parameters of the cloning state. We shall
do this by optimizing the information gained by Eve. We shall
impose that, in virtue of the ``mirror'' property of the cloning
transformation, when the detector associated to the projector
onto the state $\ket{k_{\phi^*}}_E\ket{l_{\phi}}_{E'}$ fires, the
probability of the inference that Alice and Bob's state is
$\ket{k_{\phi^*}}_A\ket{l_{\phi}}_B$ is maximal. It is easy to
check that the probability that Alice and Bob's state is
$\ket{k'_{\phi^*}}_A\ket{l'_{\phi}}_B$ conditioned on the
observation by Eve of the state
$\ket{k_{\phi^*}}_E\ket{l_{\phi}}_{E'}$ is equal to $
\displaystyle \delta_{k'-l',k-l}{ |\sum^{N-1
}_{n=0}a_{l-k,n}exp(i{2\pi\over N} (k'-k)n)|^2\over N\sum^{N-1
}_{n=0}| a_{l-k,n}|^2 }$.
Let us first assume that the fidelity and the disturbances are
fixed,(including the coefficients $a_{i0}$ ($i=0,...N-1)$). These
parameters will be varied them later. Using the method of
Lagrange's multipliers with the constraint that $\sum^{N-1
}_{j=0}| a_{ij}|^2$ is constant, and maximizing the function
$|\sum^{N-1 }_{j=0}a_{ij}|^2$ under the variations of $a_{i,1},
a_{i,2},...a_{i,N-1}$, we get that the $N-1$ dimensional vector
$(\sum^{N-1 }_{j=0}a_{ij},\sum^{N-1 }_{j=0}a_{ij},...,\sum^{N-1
}_{j=0}a_{ij})$ is parallel to the $N-1$ dimensional vector
$(a_{i,1},a_{i,2},...,a_{i,N-1})$. The solution of these
constraints that corresponds to a maximum is the following:
$a_{i,1}=a_{i,2}=...=a_{i,N-1}$ and the phase of these complex
numbers is the same as the phase of $a_{i,0}$. Now, as
$(a_{i,1},a_{i,2},...,a_{i,N-1})=(a_{0,1},a_{0,2},...,a_{0,N-1})$=$(y_1,y_2,...,y_{N-1})$
and $(a_{0,0},a_{1,0},...,a_{N-1,0})=(v,x_1,...,x_{N-1})$ in
virtue of Eq.~(\ref{asym1}), we get that $y_1=y_2=...=y_{N-1}$,
and all the coefficients $x_i$ and $y_i$ have the same phase as
$v$. We can without loss of generality assume that this phase is
zero. Moreover, we must impose that all disturbances are equal;
otherwise, Eve's presence could be detected easily by Alice and
Bob.
Indeed, at this level all the states of a same $\phi$ basis are
not treated on the same footing. This can be checked for instance
by estimating the disturbances. There are {\em N-1} possible
errors when copying the basis state $\ket{k_{\phi}}$ Depending on
it being transformed into $\ket{(k+i)_{\phi}} (\mod N$, with
$i=1,...,N-1$). Therefore, we define {\em N-1} disturbances
$D_1$, $D_2$and $D_{N-1}$ corresponding to these {\em N-1}
errors. By a straightforward but lengthy computation, we get that
the $i$th disturbance is equal to $|x_i|^2+\sum^{N-1 }_{j=0
}|y_j|^2$ ($j=1,...,N-1$) which, in general, is not independent
on the label $i$. If we impose that the disturbances are
independent of the label $i$, we get $x_1=x_2=...=x_{N-1}$ and
the matrix $a_{mn}$ contains only real positive coefficients.
Taking account of Eq.~(\ref{asym1}), we obtain the final form of
the matrix $a_{m,n}$:
\begin{equation}\label{asym2}
(a_{m,n})= \left(\begin{array}{ccccc}
v & y & y &...&y \\
x & y & y & ...&y \\
x & y  & y&...&y  \\
. & .  & .&...&.  \\
. & .  & .&...&.  \\
. & .  & .&...&.  \\
x & y  & y&...&y  \\
\end{array}\right)
\end{equation}

Note that it is sufficient that the cloning state
$\ket{\Psi}_{R,A,B,C}$ is invariant for two distinct values of
$\phi$ in order that it is invariant for all values of $\phi$, so
to say that it acts identically on each state of the $\phi$
bases. Such a cloner is thus phase-covariant, in analogy with the
qubit\cite{bruss} and qutrit cases\cite{Durttrits}.
 
We determined numerically the values of $v$, $x$ and $y$ for
which Eve's information is maximal while the fidelity of Bob's
clone is fixed. Letting vary this fidelity, we determined the
error rate that corresponds to the crossing point of Bob and
Eve's mutual information relative to Alice's data
($I_{AB}=I_{AE}$). According to Csisz\'{a}r and K\"{o}rner
theorem \cite{Csiszar} Alice and Bob can distill a secure
cryptographic key if the mutual information between Alice and Bob
$I_{AB}$ is larger than the mutual information between Alice and
Eve $I_{AE}$, i.e., $I_{AB}>I_{AE}$. If we restrict ourselves to
one-way communication on the classical channel, this actually is
also a necessary condition. Consequently, the quantum
cryptographic protocol above ceases to generate secret key bits
precisely at the point where Eve's information matches Bob's
information.
 
The threshold fidelities below which the security of the protocol
is no longer guaranteed are listed in function of the dimension
$N$ in the table \ref{tab1}. These values are the exactly the
same values as those obtained from a higher dimensional
generalization of Ref. \cite{chenzuko}.
\begin{table}
\begin{tabular}{c|c}
\hline $N$ & $F_A$ \\
\hline 2 & 0.853553 \\
3 & 0.775276 \\
4 & 0.734178 \\
5 & 0.708043 \\
6 & 0.689788 \\
7 & 0.676230\\
8 & 0.665708 \\
9 & 0.657267 \\
10 & 0.650319 \\
$\infty$ & 0.5 \\ \hline
\end{tabular}
\caption{Fidelity for dimension $2 \leq N \leq 10$}\label{tab1}
\end{table}
Note that in the qubit ($N$=2) and qutrit ($N$=3) cases, we
recover the properties (optimal fidelity, upper bound on the
error rate and so on) derived in the literature following Cerf's
approach\cite{boure1,Cloning-a-qutrit,Durttrits} or more general
approaches\cite{bruss,HBech-NGisin,dagomir}. The threshold
fidelities that we obtained are lower than the corresponding
values in the case of symmetric cloners, which were derived in
Ref.\cite{fan}. This is due to the fact that in our approach Eve
considers the full information contained in the clone (B) and in
the ancilla (C). In the large $N$ limit, it is easy to show that
$I_{AE}= 1 - F_A $ and $I_{AB} = F_A$. Thus, in this limit, we
find that the cloner converges to the universal (isotropic)
cloner while the fidelity goes to fifty percent. The tolerable
error rate has also been shown recently Ref. \cite{CHAU} to tend
to fifty percent with the dimension of qu$N$its going to infinity
(with $N$ being a prime number) in another prepare-and-measure
scheme. It seems that this asymptotic behaviour is quite general.

\subsection{Correlations between non-conjugate bases}
 
The unbiased noise that appears in Eq.~(\ref{werner}) is
characterized by a density matrix that is proportional to the
identity $\hat{I}_{R,A}$. Now, $\hat{I}_{R,A}$=
$\hat{I}_{A}$.$\hat{I}_{R}$=
$\sum_{i=0}^{N-1}\ket{\psi^*_i}_{R}\bra{\psi^*_i}_{R}$.$\sum_{j=0}^{N-1}\ket{\tilde
\psi_j}_{A}\bra{\tilde \psi_j}_{A}$ where the $\psi$ and the
$\tilde\psi$ bases can be chosen arbitrarily. This arbitrariness
suggests some invariance of the noise under local changes of
basis. In particular, when Eve replaces the signal by a clone,
this invariance must be respected. According to the condition
(3), this clone must mimic the correlations between the non-fully
correlated (non-conjugate) bases that would be observed in the
presence of real, unbiased, noise. We shall now show that this is
well the case. In order to do so, let us consider the reduced
cloning state
$\rho_{R,A}=Tr_{B,C}\ket{\Psi}_{R,A,B,C}\bra{\Psi}_{R,A,B,C}$
obtained after averaging over Eve's degrees of freedom; according
to
Eqs.~(\ref{optimal},\ref{optimalconj},\ref{def1},\ref{CERF},\ref{asym2}),
and thanks to the identities

$\ket{B^{\phi}_{0^*,0}}=\ket{\phi_N^+}$, $\sum^{N-1}_{m=0,n=0}
\ket{B^{\phi}_{m^*,n}}_{R,A}\bra{B^{\phi}_{m^*,-n}}_{R,A}=\hat{I}_{R,A}$,
and $\sum_{n=0}^{N-1} e^{-2\pi i ((k-l)n/N)}=N.\delta_{k,l}$, we
get: \beqa \rho_{R,A}=
v^2\ket{B^{\phi}_{0^*,0}}_{R,A}\bra{B^{\phi}_{0^*,0}}_{R,A}+x^2\sum^{N-1}_{m=1}
\ket{B^{\phi}_{m^*,0}}_{R,A}\bra{B^{\phi}_{m^*,0}}_{R,A}+y^2\sum^{N-1}_{m=1,n=0}
\ket{B^{\phi}_{m^*,n}}_{R,A}\bra{B^{\phi}_{m^*,-n}}_{R,A}\nonumber \\
=
(v^2-x^2)\ket{\phi_N^+}_{R,A}\bra{\phi_N^+}_{R,A}+(x^2-y^2)\sum^{N-1}_{m=0}
\ket{B^{\phi}_{m^*,0}}_{R,A}\bra{B^{\phi}_{m^*,0}}_{R,A}+y^2\sum^{N-1}_{m=0,n=0}
\ket{B^{\phi}_{m^*,n}}_{R,A}\bra{B^{\phi}_{m^*,-n}}_{R,A}\nonumber\\
=(v^2-x^2)\ket{\phi_N^+}_{R,A}\bra{\phi_N^+}_{R,A}+N.(x^2-y^2)\frac{1}{N}\sum^{N-1}_{n=0}\ket{n}_R\ket{n}_A\bra{n}_R
\bra{n}_A+N^2.y^2\rho^{noise}_{R,A}\label{reduced}\eeqa

In comparison to Eq.~(\ref{werner}) a new factor weighted by
$N.(x^2-y^2)$ appears in the previous expression. It is a mixture
of projectors on products of the states of the computational
basis $\ket{n}_R\ket{n}_ A$.
 
For what concerns measurements performed by Alice and Bob in the
optimal bases (that they are conjugate or not), any such product
has the same statistical properties as the unbiased noise
$\rho^{noise}_{R,A}=\frac{1}{N^2}\hat{I}_{R,A}$. The deep reason
for this property is that the in-product between any state of the
computational basis and any $\phi$ state is in modulus squared
equal to ${1\over N}$, in other words, both bases are mutually
unbiased. Henceforth, the modulus squared of
$\bra{n}_R\bra{n}_A\ket{k_{\phi_1}^*}\ket{l_{\phi_2}}$ is equal
to ${1\over N^2}$, whatever the values of the labels $k$ and $l$
could be.

\section{Conclusions}
 
The Ekert91 protocol\cite{Ekert} and its qu$N$it extension, the
$N$-DEB protocol which is analyzed in the present paper, involve
encryption bases for which the violation of local realism is
maximal. If Alice and Bob measure their member of a
maximally-entangled qu$N$it pair in two ``conjugate'' bases, this
gives rise to perfect correlations. After  measurement is
performed on each member of a sequence of maximally-entangled
qu$N$it pairs, Alice and Bob can reveal on a public channel what
were their respective choices of basis and identify which qu$N$it
was correctly distributed, from which they will make the key.
They can use the rest of the data in order to check that it does
not admit a local realistic simulation. For instance they can
check that their correlations violate some generalized Bell or
CHSH inequalities\cite{collins}. More generally, they can check
that the correlations between their results (that they are
perfectly correlated or not are the same as the results that they
expect in the presence of unbiased noise. Note that the optimal
bases do not allow them to differentiate a fully unbiased noise
(described by a fully incoherent reduced density matrix
proportional to the unity matrix) from a ``colored'' noise that
would contain projectors (with arbitrary weights) on product
states of the computational basis. As shown at the end of the
last section, this is due to the fact that all the optimal bases
are mutually unbiased relatively to the computational basis. This
explains why the phase-covariant attacks are more dangerous than
the universal (state-independent) attacks. For instance, the
maximal admissible error rate (when attacks based on
state-dependent cloners are considered) was shown to be equal to
$E_A=1-F_A=1-({1\over 2}+{1\over\sqrt{8}}) \simeq 14.64\%$ for
the 2-DEB protocol (or Ekert91 qubit protocol)
\cite{bruss,HBech-NGisin,Cloning-a-qutrit} and to 22.47 \% for
the 3-DEB protocol\cite{dagomir,Durttrits}. The corresponding
rates, if we restrict ourselves to state-independent
cloners\cite{CERF} were shown in Ref.\cite{boure1} to be
respectively equal to 15.64\% and 22.67\%. These results were
derived for a slightly asymmetric state-independent
cloner\cite{CERF,boure1} that clones all the states with the same
fidelity. Universal attacks correspond to protocols in which
Alice and Bob have the physical possibility to measure
(distinguish) experimentally any coefficient of the reduced
density matrix which is not the case here. This is the price to
pay, but, at the same time, as the resistance of the violation of
local realism against noise is maximal when the
maximally-entangled qu$N$it pair is measured in the optimal
qu$N$it bases discussed here\cite{zukozeil,Durt}, the $N$-DEB
protocol is optimal from the point of view of the survival of
non-classical correlations in a noisy environment.
 
Actually, it has been shown\cite{collins} that the violation of a
Bell inequality extended to qu$N$its is possible, as long as the
``visibility" of the two-qu$N$its interference exceeds a
threshhold value $V_{thr}$ given by the equation ${N^2\over
V_{thr}(N)}$ = $\sum_{k=0}^{[N/2]-1}(1-{2k\over N-1})({1\over
sin^2(\pi (4k+1)/4N)}-{1\over sin^2(\pi (4k+3)/4N)})$.
The visibility mentioned above is directly related to the
threshold fraction of unbiased noise, $(1-V_{thr}(N))$, which has
to be admixed to the maximally entangled state in order to erase
the non-classical character of the correlations, and therefore is
a measure of robustness of non-classicality (see also
\cite{zukozeil,Durt}).
This means that the non-existence of a local realistic model of
the correlations is guaranteed if the fidelity $F_{thr}$ that
characterizes the communication channel between Alice and Bob,
(detectors included, so 1-$F_{thr}$ is the effective error rate
in the transmission) is larger than ${N-1\over N}\times
V_{thr}(N) +{1\over N}$. For instance, for $N=2,3,4,5$ and 10,
1-$F_{thr}$ is equal to 14.64 \%, 20.26 \%, 23.21 \%, 25.03 \%
and 28.77 \% respectively. On the other hand, the $N$-DEB
protocol is secure against a cloning-based individual attack, if
$F=1-E_A > F_A = F_{thr}$ (see table). If we compare the previous
values of 1-$F_{thr}$ with the corresponding values of 1-$F_{A}$
(with $F_A$ given in the table), it is easy to check that when a
violation of local realism occurs, the security of the $N$-DEB
protocol against individual attacks is automatically guaranteed.
Therefore, the violation of Bell inequalities is a {\em
sufficient} condition for security, as it implies that Bob's
fidelity is higher than the security threshold. For qubits the
sufficient condition ($F_A
> 0.8436$) is also necessary (\cite{RMP}).
\par

In addition, the violation of Bell inequalities guarantees that
the $N$-DEB protocol is secure against so-called Trojan horse
attacks during which the eavesdropper would control the whole
transmission line and replace the signal by a fake, predetermined
local-variable dependent, signal that mimics the quantum
correlations (see Ref.\cite{Durttrits} and references therein).
All the protocols in which no entanglement is present (such as
BB84\cite{BB84}, the 6-state qubit
protocol\cite{HBech-NGisin,DBruss}, or the 12-state qutrit
protocol\cite{HBech-APeres}) admit a local realistic model, so
that they are not secure against Trojan horse attacks, although,
according to the results of Ref.\cite{boure1} they are slightly
more resistant against noise than the $N$-DEB protocol.
 
\par
 
As we have already noted, our results confirm the results that can
be found in the literature relatively to the security of the 2-DEB
and 3-DEB protocols, but at the same time our results are
confirmed by the corresponding results in the case that they were
derived under constraints more general than the ones that we
postulated. Note that in order to find an expression for the
cloning state that is valid for arbitrary dimension, it is
impossible presently to avoid some extra-assumptions, for
instance that the optimal cloning state is pure, $N^4$
dimensional, symmetric under cyclic permutations of the labels of the
optimal bases and so on. These assumptions are very reasonable
anyhow. If we would try to avoid any extra-constraint, the
complexity of the problem would increase with the dimension $N$
and we could not find a solution for all values of $N$. Note also
that the security of quantum cryptographic protocols against
incoherent attacks was never clearly established, simply because
the problem is too complicate to tackle.

In summary, we have established the generality of Cerf's approach
of quantum state-dependent cloning machines under fairly general
assumptions. We have shown that the acceptable error rate of the
$N$-DEB protocol turns out to increase with the dimension $N$.
Our analysis confirms a seemingly general property that the
robustness against noise of qu$N$it schemes increases with the
dimensionality $N$.
\par

\medskip

\leftline{\large \bf Acknowledgment}
\medskip One of the authors (TD) is a
Postdoctoral Fellow of the Fonds voor Wetenschappelijke Onderzoek,
Vlaanderen. TD would like to thank National University of
Singapore for royal hospitality and acknowledge support from the
IAUP programme of the Belgian government, and the grant V-18.

\section{Appendix: Invariance of the cloning state.}
Let us consider a reference basis, the $\psi$ basis, its
conjugate basis the $\psi^*$ basis, another basis, the
$\tilde\psi$ basis and its conjugate basis the $\tilde\psi^*$
basis (with $\braket{i}{\psi_j}$ = $ U_{ ij}$ and
$\braket{i}{\tilde\psi_j}$ = $\tilde U_{ ij}$). Let us assume
that during the realisation of a quantum cryptographic protocol,
Alice and Bob share the maximally entangled state
$\ket{\phi_N^+}$ ($\ket{B_{0,0}}$)  and that Alice measures it
either in the $\psi^*$ basis or in  the $\tilde\psi^*$ basis
(with $\braket{i}{\psi^*_j}$ = $ U^*_{ ij}$ and
$\braket{i}{\tilde\psi^*_j}$ = $\tilde U^*_{ ij}$). according to
the proof given in the section 2, she projects Bob's component
onto either the $\psi$ or the $\tilde\psi$ basis. If we require
that the cloning machine is invariant in both bases, the joint
(Cerf) state of the reference $R$, the two output clones ($A$ and
$B$), and the ($N$-dimensional) cloning machine $C$
(Eq.~(\ref{CERF})) must fulfill the following condition:

\beq \sum_{m,n=0}^{N-1} a_{m,n} \; \ket{ B_{m^*,n}}_{R,A} \ket{
B_{m,-n^*}}_{B,C} = \sum_{m,n=0}^{N-1} a_{m,n} \; \ket{\tilde
B_{m^*,n}}_{R,A} \ket{\tilde B_{m,-n^*}}_{B,C} \eeq
 
As the $N^2$ $B_{m,n}$ states form an orthonormal basis, we can
project the righthand side of the previous equality onto them,
which gives: $\sum_{m,n=0}^{N-1} a_{m,n} \ket{B_{m^*,n}}_{R,A}
\ket{B_{m,-n^*}}_{B,C} = \sum_{m,n,m',n',m'',n''=0}^{N-1} a_{m,n}$
 $\ket{B_{m'^*,n'}}_{R,A}\  _{R,A}\braket{B_{m'^*,n'}}{\tilde B_{m^*,n}}_{R,A}$
$\ket{B_{m'',-n''^*}}_{B,C} $$ _{B,C}\braket{B_{m'',-n''^*}}{\tilde B_{m,-n^*}}_{B,C}$ 

Denoting
$V_{i,j,k,l}$ the in-product $\braket{B_{i^*,j}}{\tilde
B_{k^*,l}}$, we get:
$\sum_{m,n,k,l=0}^{N-1} a_{m,n}
\delta_{(m,n),(k,l)}\ket{B_{m^*,n}}_{R,A} \ket{B_{k,-l^*}}_{B,C} =$

$\sum_{m,n,m',n',m'',n''=0}^{N-1} a_{m,n}$
 $\ket{B_{m'^*,n'}}_{R,A}  V_{m',n',m,n}$
$\ket{B_{m'',-n''^*}}_{B,C} V^*_{m'',n'',m,n}$
=$\sum_{m,n,i,j,k,l=0}^{N-1} a_{i,j}\delta_{(i,j),(i',j')}$
$V_{m,n,i,j}$ $V^*_{k,l,i',j'}$
 $\ket{B_{m^*,n}}_{R,A}  \ket{B_{k,-l^*}}_{B,C} $.
Thanks to the orthonormality of the Bell bases, this constraint
can be expressed as a matrix relation of the form ${\cal V}{\cal
A}={\cal A}{\cal V}$ where ${\cal V}$ and ${\cal A}$ are
$N^2$x$N^2$ matrices defined as follows: ${\cal V}_{i,j;k,l}$ =
$\braket{B_{i^*,j}}{\tilde B_{k^*,l}}$ and ${\cal A}_{i,j;k,l}=
a_{i,j}\delta_{(i,j),(k,l)}$. Such a system of linear equations
is extremely simple to solve: if ${\cal V}_{i,j;k,l}\not = 0$,
then $a_{i,j}$ = $a_{k,l}$. The procedure to follow in order to
build a cloning machine that is invariant in the $\psi$ basis and
the $\tilde\psi$ basis is thus straightforward: compute the $N^4$
in-products ${\cal V}_{i,j;k,l} = \braket{B_{i^*,j}}{\tilde
B_{k^*,l}}$ ($i,j,k,l=0,...,N-1$); if ${\cal V}_{i,j;k,l}\not =
0$, then $a_{i,j}$ = $a_{k,l}$. The solutions $a_{m,n}$ of this
set of equations define the most general Cerf state
(Eq.~(\ref{CERF})) that is invariant in the two bases.

\end{document}